\documentclass[5p,authoryear,10pt]{elsarticle}
\usepackage{amssymb}
\usepackage{graphicx}
\usepackage{natbib}
\usepackage{color}

\journal{New Astronomy}

\begin{document}
\bibliographystyle{elsarticle-harv}
\begin{frontmatter}

\title{Star cluster evolution in dark matter dominated galaxies}
\author[nott,swin,kap]{Anneke Praagman\corref{cor}}
\cortext[cor]{Corresponding author.}
\ead{a.k.praagman@alumnus.rug.nl}
\author[swin]{Jarrod Hurley}
\ead{jhurley@astro.swin.edu.au}
\author[lei]{Chris Power}
\ead{chris.power@astro.le.ac.uk}

\address[nott]{School of Physics and Astronomy, University of Nottingham,
  Nottingham, NG7 2RD, UK}
\address[swin]{Centre for Astrophysics and Supercomputing, Swinburne
  University of Technology, PO Box 218, VIC 3122, Australia}
\address[kap]{Kapteyn Astronomical Institute, University of Groningen, Postbus
  800, NL-9700 AV Groningen, the Netherlands}
\address[lei]{Department of Physics \& Astronomy, University of Leicester,
  University Road, Leicester LE1 7RH, UK}

\begin{abstract}
We investigate the influence of the external tidal field of a dark matter halo
on the dynamical evolution of star clusters using direct $N$-body simulations, 
where we assume that the halo is described by a Navarro, Frenk \& White
mass profile which has an inner density cusp. We assess how varying the mass
and concentration of the halo affects the rate at which the star cluster loses 
mass and we find that increasing halo mass and concentration drives enhanced 
mass loss rates and in principle shorter cluster disruption timescales. In 
addition, we examine disruption timescales in a three-component model of a 
galaxy (bulge, disk and dark matter halo) and find good agreement with results 
based on an empirical model of the Galactic potential if we assume a halo mass 
of $\sim 10^{12}\,\rm M_{\odot}$. In general, dark matter halos are expected
to contribute significantly to the masses of galaxies and should not be
ignored when modelling the evolution of star clusters. We extend our results 
to discuss how this can have a potentially profound effect on the disruption 
timescales of globular clusters, suggesting that we may underestimate the rate 
at which primordial globular clusters are disrupted. 
\end{abstract}


\end{frontmatter}
\section{Introduction}
\label{intro}

Globular clusters are compact stellar systems, each containing of order 
$10^6$ stars, that orbit around the centres of galaxies out to large radii. 
Our Galaxy hosts around 200 globular 
clusters and observations show that almost all galaxies host these systems, 
with giant ellipticals having the largest (relative) population 
\citep{brodie2006}. 
Stellar population studies have revealed that globular clusters have ages up to
$\sim 13\,$Gyrs \citep[e.g.][]{hansen2002,chaboyer2002}. This implies 
that they formed within $\sim 1$ billion years of the Big Bang and so they 
represent fossil records of the earliest epoch of galaxy formation, which has
led to much interest in their potential as probes of the high redshift Universe
\citep{brodie2006}. 

However, we have as yet no compelling theory for globular cluster
formation and evolution within a cosmological framework and this has important 
consequences for the kind of questions we can address using globular clusters 
and the strength of conclusions that we can draw. One particularly important 
consequence concerns the disruption of primordial globular clusters and the
relation of the high redshift globular cluster population to the population that we
observe at the present day. All star clusters lose mass over
time and this depends on a number of factors including initial conditions
(e.g. initial mass profile of the star cluster, initial mass function of the 
stars, initial 
binary fraction), internal processes within the cluster (e.g. mass segregation, 
stellar mass loss), the orbital parameters of the cluster and the form and
nature of the external tidal field within which the cluster orbits. The
globular cluster population that we observe around galaxies at the present day
has survived a Hubble time, but it is interesting to ask whether or not this
population is representative of the primordial globular cluster population at high
redshift. If not, this would imply that many primordial globular clusters 
disrupted on relatively short timescales, and that the clusters that survived
to the present day are in a sense atypical. This introduces certain
caveats where the present day population is used to probe, 
say, the efficiency of star formation at high redshifts \citep{spitler2008} and the 
epoch of cosmological reionisation \citep{moore2006}.

In the standard picture of galaxy formation, galaxies are embedded in massive 
virialised structures or halos of dark matter \citep{springel2006}. These 
dark matter halos assemble over time hierarchically, continuously growing via 
accretion of dark matter and merging with other halos from high redshift to
the present day. The subset of globular clusters that have survived to the
present day has evolved in a time-dependent and at times violently changing
potential provided by its host galaxy and dark matter halo. 
Previous studies that have investigated star cluster mass loss 
\citep[e.g.][]{giersz1997, vesperini1998, baumgardt2001, hurley2001} 
-- with models ranging in mass from open clusters to small globular clusters 
-- have assumed simplistic models for the Galaxy's tidal field by 
treating the Galaxy as a central compact source or by using Oort constants that
are empirically measured from stellar motions in the Solar neighbourhood 
\citep[e.g.][]{bandt}. \citet{mashchenko2005} looked at the early evolution of 
globular clusters formed in dark matter halos but employed softened 
gravitational potentials 
for both the stars and dark matter. 
As such, the explicit presence\footnote{We use explicit because models of the 
Galactic potential based on Oort constants implicitly include the contribution 
of the Galaxy's dark matter halo.} of a dark matter halo has been neglected to 
date in direct $N$-body models of star cluster evolution. 

Our goal is to understand how the presence of a dark matter halo affects the 
internal characteristics of star clusters in general -- and globular clusters in
particular -- and their disruption rates. For the 
former our interest is in signatures such as the ratio of the core to
half-mass radii and the velocity dispersion, which may be compared to 
observations of star clusters. However, in this paper, the first in a series, 
we start by focusing on the rate of mass-loss from model star clusters and 
what this can tell us about the expected lifetimes and disruption of globular 
clusters. Our approach is to use a direct $N$-body code that treats the 
internal processes of a star cluster in as detailed a manner as possible. We 
include a spectrum of stellar masses and stellar evolution from the outset as 
the effect of these on star cluster evolution is relatively well understood 
\citep[e.g.][]{dlfm1996, baumgardt2003, hurley2004}. In our treatment of the 
external potential and the orbits of the model clusters within this, we take a 
more cautious approach. We start with the simplest case, circular orbits in a 
variety of static potentials, as this facilitates comparison with previous 
$N$-body studies and provides the foundation for our study. In the future we 
will expand the models to include eccentric orbits and time-varying
potentials. By necessity we also start with models more comparable in size to 
open clusters than globular clusters but this will also increase as we 
progress. In this way we build a consistent and realistic picture of star 
cluster evolution and survival.

We assess 
the influence of halo structure on mass loss rate using the functional form
for the mass profile proposed by \citet*[hereafter NFW]{nfw1996,nfw1997}. The
NFW profile has been found to provide a good description of the spherically 
averaged dark matter distribution within halos in dynamical equilibrium in 
cosmological $N$-body simulations. The characteristic feature of the NFW
profile is that it has a central density cusp, $\rho \propto r^{-1}$, so that
the central density is divergent. We also construct a three-component model of 
our Galaxy consisting of a bulge, disk and NFW halo and we compare the rate at 
which clusters lose mass in this model against the rate measured in the
fiducial model of the Galactic tidal field, based on Oort constants.

The format of the paper is as follows. In the next section we describe the 
NFW profile in more detail, while in
section~\ref{cluster_model} we describe the model we have adopted for our star
cluster. In section~\ref{results} we present the main results of our study,
first considering the evolution of an isolated cluster; then the evolution of
the cluster in external galactic potentials; then evolution in the potential of
a NFW halo; and finally evolution in our three-component model of a galactic 
potential, which consists of a bulge, disk and NFW halo. Finally we discuss
the main findings and implications of our study in section~\ref{summary}.

\section{The NFW Mass Profile}
\label{nfw_profile}

Dark matter halos that form in cosmological simulations are relatively 
complex structures - they are generally aspherical \citep[e.g.][]{bailin2005}
and asymmetric \citep[e.g.][]{gao2006}, they have no simple boundary 
\citep[e.g.][]{prada2006}, and they contain a wealth of small scale structure 
\citep{gao2004}. Despite this complexity, however, it is conventional to identify a
halo as a spherical overdense region, typically of order 100 times the mean
density of the Universe. The mass enclosed within this spherical overdensity
defines the \emph{virial mass} of the halo,

\begin{equation}
  \label{eq:mvir}
        {M_{\rm vir}=\frac{4\pi}{3} \Delta_{\rm vir} \rho_{\rm crit} r_{\rm vir}^3.}
\end{equation}

\noindent Here $\rho_{\rm crit}=3H^2/8\pi G$ is the critical density of the 
Universe and $r_{\rm vir}$ is the \emph{virial radius}, which defines the radial 
extent of the halo. $\Delta_{\rm vir}$, the virial overdensity criterion, 
corresponds to the mean overdensity at the time of virialisation in the 
spherical collapse model, the simplest analytic model of halo formation  
\citep[cf.][]{eke1996}. Depending on redshift and cosmological parameters, 
$\Delta_{\rm vir}$ varies between  $\sim 100$ and $\sim 200$. 

The mass profiles of dark matter halos in dynamical equilibrium forming 
in cosmological $N$-body simulations can be relatively well described by 
the NFW profile,
\begin{equation}
  \label{eq:nfwrho}
  {\rho(r) = \frac{\rho_{\rm crit} \delta_{\rm c}}{r/r_{\rm s}\,(1+r/r_{\rm s})^2},}
\end{equation}

\noindent where $\rho_{\rm crit}$ is the critical density of the Universe,
$r_{\rm s}$ is the scale radius and $\delta_{\rm c}$ is the characteristic overdensity.
The characteristic overdensity $\delta_{\rm c}$ is itself a function of $r_{\rm s}$,
\begin{equation}
\label{eq:deltac}
{\delta_{\rm c} = \frac{\Delta_{\rm vir}}{3}\frac{c^3}{\ln (1+c)-c/(1+c)},}
\end{equation}
\noindent where $c=r_{\rm vir}/r_{\rm s}$ is the concentration. 

Because the scale radius $r_{\rm s}$ and the characteristic overdensity $\delta_{\rm c}$ are 
related, equation~\ref{eq:nfwrho} can be rewritten in terms of a single 
parameter, the concentration $c=r_{\rm vir}/r_{\rm s}$, such that for a fixed 
concentration, the local density depends only on the normalised radius 
$r/r_{\rm vir}$. Cosmological simulations have shown that virial mass 
$M_{\rm vir}$ and concentration $c$ are correlated, such that the 
concentration increases as the virial mass decreases
\citep[e.g][]{bullock2001,eke2001,neto2007}. As a rule of thumb, a typical 
$M_{\rm vir}$=$10^{10}$ ($10^{12},10^{14}$) $\rm M_{\odot} $ halo has a
concentration $c \simeq 15$ (10,5).

The mass enclosed within a radius $r$ is obtained from equation~\ref{eq:nfwrho};
\begin{equation}
  \label{eq:nfwmass}
        {M(<r) = 4 \pi \rho_{\rm crit} \delta_{\rm c} r_{\rm s}^3 g(r)}
\end{equation}
\noindent where 
\begin{equation}
  \label{eq:gofr}
        {g(r) = \ln(1+r/r_{\rm s})-\frac{r/r_{\rm s}}{1+r/r_{\rm s}}.}
\end{equation}
\noindent This results in a radial acceleration directed towards the centre of the
halo of
\begin{equation}
  \label{eq:nfwacc}
        {a(r) = -4 \pi G \rho_{\rm crit} \delta_{\rm c} r_{\rm s} g(r) \left(r/r_{\rm s}\right)^{-2}.}
\end{equation}
\noindent It follows that the gradient in the radial acceleration (and consequently the force)
is given by
\begin{equation}
  \label{eq:nfwdadr}
        {\frac{\partial a}{\partial r} = -2\frac{a(r)}{r}-4\pi G\rho(r).}
\end{equation}
This provides a measure of the tidal force across the cluster and in the absence of an 
extended dark matter halo the tidal force would reduce to the familiar result for a point mass, 
\begin{equation}
\frac{\partial a}{\partial r} = -2\frac{GM(r)}{r^3}
\end{equation}

\section{The cluster model}
\label{cluster_model}

\begin{table}
\begin{center}
\begin{tabular}{cl}
Model & Description \\
\\
1& Isolated cluster\\
2& Standard Galactic tidal field\\ 
3& Two-component Galaxy (disk and bulge)\\
4& NFW halo\\
5& Full Galaxy (disk, bulge and NFW halo)\\
\end{tabular}
\caption{Overview of the models used in the simulations in terms of the
  treatment of the external tidal field.} \label{tb:models}
\end{center}
\end{table}

To study the evolution of a star cluster we used the $N$-body code {\small
  NBODY6} \citep[see][for a full description]{aarseth2003}. It uses Hermite 
integration with individual time-steps and does not use softening in the 
force equation. Stellar evolution is included using the algorithms provided 
by \citet{hurley2000} as described in \citet{hurley2008}. The code also 
allows for binary formation and evolution. There are also options to include 
an external Galactic potential with the simplest option being a standard 
Galactic potential at $8.5\,$kpc based on the local Oort constants (hereafter
the standard Galactic tidal field). In most respects {\small NBODY6} is 
very similar to its sister code {\small NBODY4}
\citep{aarseth1999,hurley2001} except that the latter interfaces
with special-purpose GRAPE hardware \citep{makino2003} to speed up the
calculation of the gravitational forces between the stars. {\small NBODY6}
operates on standard hardware and our simulations are performed on
the Swinburne supercomputer\footnote{see
http://astronomy.swin.edu.au/supercomputing/}.

Within {\small NBODY6} there is the possibility of building a multi-component 
galaxy consisting of bulge (point mass), disk \citep{miya1975} and logarithmic 
potential \citep{aarseth2003}. To this we have added the possibility of
replacing the logarithmic potential with a potential based on the
NFW-profile. This three-component galactic potential thus consists of a bulge, 
disk and NFW halo. In all other respects the {\small NBODY6} software is the 
same as the version publicly available for download at 
{\tt http://www.ast.cam.ac.uk/~sverre/web/pages/nbody.htm} with input options 
and parameters set as for the defaults suggested in the included manual and 
described by \citet{aarseth2003}.

All our models start with single stars only and these are distributed
according to the \citet{plummer1911} density model, an $n=5$ polytrope.
We assume the stars are in virial equilibrium when setting the initial 
positions and velocities and take the initial virial radius to be $3\,$pc 
which sets the length scale of the simulation. It should also be noted that 
the Plummer profile formally extends to infinite radius so a cut-off at 10 
times the half-mass radius is applied to deal with rare cases of large
distance \citep{aarseth2003}. To save computing time we did not run full 
globular cluster models (i.e. $N=10^5-10^6$). Instead we focus on models of 
$1 000$ stars. For one galaxy scenario we also ran $N=16\,000$ 
simulations in order to investigate the scaling of the results with
$N$. The initial masses of the stars are distributed according to a Salpeter
initial mass function (IMF: $n(M)\propto M^{-2.3}$) with a minimum and maximum
stellar mass $m_{\rm min} = 0.3\,{\rm M_{\odot}}$ and $m_{\rm max}= 30\,{\rm
  M_{\odot}}$ respectively. The average stellar mass is $1\,{\rm
  M_{\odot}}$.
In general our clusters are placed on circular orbits in the 
stellar disk (if present) at a radius of $8.5\,$kpc from the Galactic centre. 
This facilitates comparison with previous studies. However, we do briefly 
explore the effect of orbital distance by placing some clusters on orbits at 
radii of 4.5, 17 and $34\,$kpc. An overview of our models used in terms of the
treatment of the external tidal field is shown in Table \ref{tb:models}.

\section{Results}
\label{results}
\subsection{Star Cluster Evolution in Isolation}
\begin{figure}
\begin{center}
\includegraphics[width=9cm]{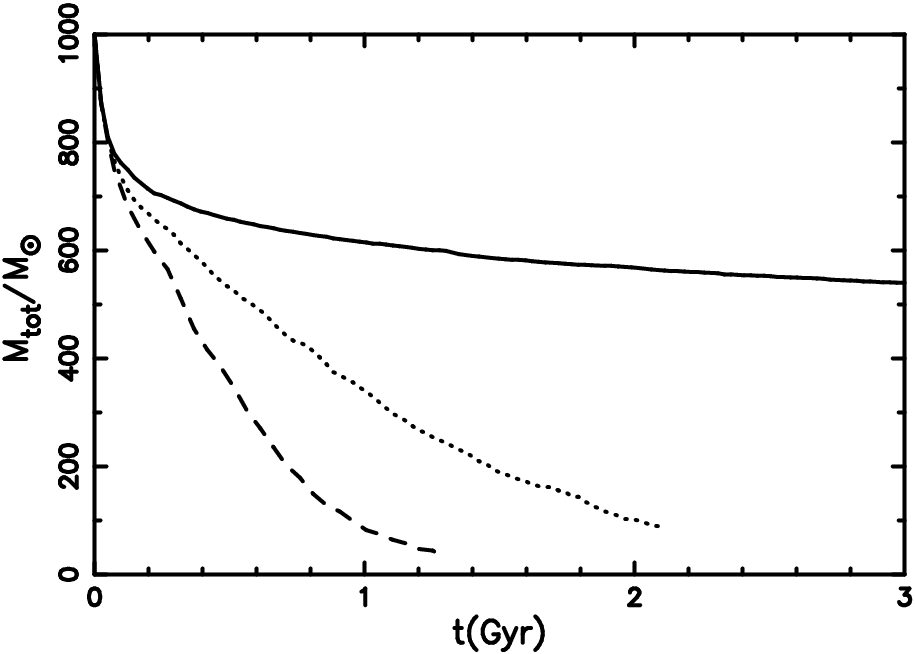}
\caption{The evolution of the total mass of the cluster in isolation (Model 1,
  solid), in the standard Galactic tidal field (Model 2, dashed) and in the
  two-component galaxy (Model 3, dotted). All models start with $N=1000$.} 
\label{fig:mass123}
\end{center}
\end{figure}

Our first model (Model 1) is a cluster in isolation. Its dynamical evolution
depends only on internal processes such as mass segregation and stellar 
evolution. The evolution of the total mass of such a cluster is shown in
Figure \ref{fig:mass123} by the heavy solid curve. In {\small NBODY6} the 
cluster size is limited by a tidal radius or cut-off radius. For the case of
an isolated cluster it is not meaningful to define a tidal radius and so we
adopt a cut-off radius of 10 times the cluster scale length $r_{\rm s}$,
\begin{equation}
\label{eq:r_s}
r_{\rm s} = \frac{1}{2}\frac{GM_{\rm tot}^2}{\Phi},
\end{equation}
with $M_{\rm tot}$ the total cluster mass and $\Phi$ the potential energy. 
Note that $r_{\rm s}$ is distinct from the NFW scale radius discussed in 
section~\ref{nfw_profile}. Stars with cluster-centric radii in excess of twice the 
cut-off radius are considered to have left the cluster and contribute to the 
mass loss. The initial cut-off radius is $23\,$pc, but this increases as the 
cluster expands to about $100\,$pc after $3\,$Gyr. 

Mass segregation causes the system to lose stars early on. Less massive stars 
gain kinetic energy through equipartition during encounters with more massive 
stars in the core and are more likely to escape. Stellar evolution leads to 
the loss of even more mass in the early stages. Massive stars evolve more 
rapidly than low mass stars and hence lose mass more quickly in the form of 
stellar winds. Neutron stars that form in a supernova will get a velocity kick 
and generally disappear from the system.  

\subsection{The Standard Galactic Tidal Field Model}

Model 2 introduces the standard Galactic tidal field (SGTF) based on the local Oort constants ($A = 14.4$ km/s, $B = -12.0$ km/s). 
These are derived empirically from the observed motions of
stars in the Solar neighbourhood \citep{bandt}. As such they naturally account for the local 
effect of the total mass of the Galaxy (including its dark matter halo) and
they strictly apply to orbits within 
the region of the Solar neighbourhood. From the local circular velocity we can 
infer that the enclosed mass at an orbit of $8.5\,$kpc (the distance from the 
Sun to the Galactic centre) is $9\times 10^{10}\,
{\rm M_{\odot}}$. As the dashed curve in Figure \ref{fig:mass123} shows this has a
profound effect on the mass loss of the cluster. There is an almost continuous 
mass loss leading to rapid disruption of the cluster: simulations are halted 
when the cluster has less than 25 stars left. Here we define the tidal radius 
as the \citet{king1962} tidal radius,  
\begin{equation}
\label{eq:king}
r_{\rm t}^3=\frac{1}{3}\frac{M_{\rm C}}{M_{\rm G}}R_{\rm G}^3,
\end{equation}
with $M_{\rm C}$ and $M_{\rm G}$ the masses of the cluster and the galaxy 
respectively and $R_{\rm G}$ the distance between the galactic centre and 
the centre of the cluster. We set $R_{\rm G} = 8.5\,$kpc (the distance from 
our Sun to the Galactic centre). This gives a tidal radius of $\sim 16\,$pc,
smaller than the cut-off radius used in Model 1. This has a bearing 
on what we compute as the mass loss rate, which will be lower in the case of the 
isolated cluster where the cut-off radius is higher. However, stellar
evolution effects dominate in the isolated cluster so the general comparison holds.   

\subsection{The Two-Component Model -- Bulge \& Disk}
\label{sc:diskbulge}

Model 3 introduces the two-component model of the Galaxy, consisting of a
bulge and disk. The disk is modelled as a Miyamoto-Nagai disk \citep{miya1975} 
and the bulge is a point mass at the centre of the galaxy. We use a mass of 
$5\times10^{10}\,{\rm M_{\odot}}$ for the disk, which has a radius of $40\,$kpc,
and $1.5\times10^{10}\,{\rm M_{\odot}}$ for the bulge, as suggested by 
\citet{Xue2008} for the Milky Way. 

We adopt a tidal radius of $16\,$pc, equal to the King radius (see Equation
\ref{eq:king}) of Model 2. This is kept fixed for all subsequent models at an
orbital distance of $8.5\,$kpc, including the models where we add a dark matter halo. 
Thus we are effectively employing a tidal cut-off as described by \citet{trenti2007}. 

The dotted line in Figure \ref{fig:mass123} shows the evolution of the cluster
mass with Model 3 acting as the external tidal field. Model 3 does not produce
a tidal field as strong as the SGTF of Model 2, but 
this is to be expected because the mass of the combined bulge and disk within 
$8.5\,$kpc is smaller than the enclosed mass one infers for the SGTF of Model
2, which is $9\times 10^{10}\,{\rm M_{\odot}}$. Recall that Model 2 is based on an
empirical estimate of the Galactic tidal field and so implicitly includes the 
contribution of the Galaxy's dark matter halo, whereas Model 3 has no halo contribution.

The two-component model is attractive because we can fully determine
how the Galactic potential is modelled and explore different sets of
assumptions, but it neglects the presence of dark matter halo component
to the potential which we expect to be important. In contrast, the
SGTF model provides a realistic model of the
Galactic potential in the Solar neighbourhood, being derived from
empirical measurement, but the conditions under which it is applicable
are restrictive. If the two-component model is to be made more
realistic, then it must become a three-component model (a bulge, disk
and halo), in which case we expect the mass loss rates for Models 2 and
3 to converge. In the next subsection we look at the influence of a
single component model -- that of a NFW dark matter halo on the mass
loss rate -- before introducing the three-component model in the final
subsection.

\begin{figure}
\begin{center}
\includegraphics[width=9cm]{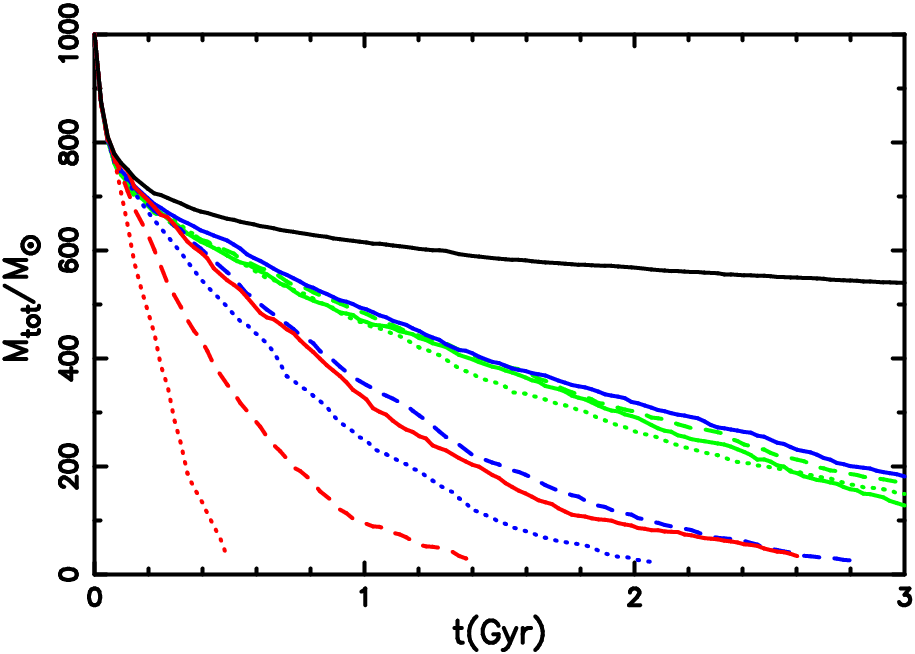}
\caption{The total mass evolution of a cluster of 1000 stars in an NFW halo at
  an orbital distance of $8.5\,$kpc from the halo centre (Model 4). We show here
  three different virial masses: $10^9\,{\rm M_{\odot}}$ (green), $1.5 \times
  10^{12}\,{\rm M_{\odot}}$ (blue) and $10^{14}\,{\rm M_{\odot}}$ (red); each
  with three different concentrations: $c = 5$ (solid), $10$ (dashed) and $15$
  (dotted). The solid black line is Model 1, the cluster in isolation, and is
  drawn here as a reference.} \label{fig:massNIII8}
\end{center}
\end{figure}

\subsection{The NFW Halo}
We now consider a set of models in which our star clusters orbit in a NFW dark
matter halo (Model 4). For now we ignore the stellar component of the galaxy
because we want to investigate the influence of changing the NFW parameters of 
mass and concentration. We use a model grid of 6 different halo virial masses 
($10^{9}\,{\rm M_{\odot}}$, $10^{10}\,{\rm M_{\odot}}$, $10^{11}\,{\rm
  M_{\odot}}$, $1.5 \times 10^{12}\,{\rm M_{\odot}}$, $10^{13}\,{\rm M_{\odot}}$
and $10^{14}\,{\rm M_{\odot}}$) and 3 concentrations (5, 10 and 15), thus 18
models in total. Here we define the virial mass as in equation~\ref{eq:mvir}
with an overdensity criterion of $\Delta_{\rm vir}=200$. 
Each cluster follows a circular orbit at a distance of $8.5\,$kpc from the
centre of the halo. This is achieved by giving the cluster an initial circular
velocity equal to the halo velocity at that distance,
\begin{equation}
V_{\rm c}=\sqrt(GM(<r)/r).
\end{equation}
Our main set of simulations start with 1000 stars with a Salpeter IMF in the
mass range $0.3 - 30.0\,{\rm M_{\odot}}$ and total mass of $1000\,{\rm M_{\odot}}$. Figure 
\ref{fig:massNIII8} shows the evolution of the mass of the cluster. For
clarity we only show 9 of the models representing the lowest mass, intermediate mass and
highest mass halos. The mass loss is initially the same for all models since in this
stage it is dominated by stellar evolution of the most massive stars. After about $100\,$Myr
clusters in the most massive halos keep losing mass at a faster pace than the 
clusters in the less massive halos. In these massive halos the effect of concentration 
is also clearly visible. Higher concentrations mean faster disruptions -- 
more of the halo mass is concentrated within the cluster's orbit resulting in
a greater tidal force on the cluster. The effects of concentration become
negligible for the least massive halos, and it is also noteworthy that mass
loss is very similar for the low- and intermediate-mass halos.

\begin{table}
\label{tab:disruption}
\begin{center}
{\small 
\begin{tabular}{lllllrr}
Mass& $c$ & $\% r_{\rm vir}$ & $V_c$ & $dV_c /dr$ & $N$ & $t_{\rm dis}$\\
 (${\rm M_{\odot}}$) &&&(km/s)&(km/s/kpc)&&({\rm Myr})\\
\\
$1.0 \times 10^{9}$ & 5 & 52 & 12.46 & 0.68 & 1000 & 853\\
$1.0 \times 10^{9}$ & 10 & 52 & 13.27 & 0.55 & 1000 & 916\\
$1.0 \times 10^{9}$ & 15 & 52 &  13.66 & 0.49 & 1000 & 851\\
\\
$1.0 \times 10^{10}$ & 5 & 24 & 17.77 & 1.27 & 1000 & 859\\
$1.0 \times 10^{10}$ & 10 & 24 & 20.79 & 1.17 & 1000 & 921\\
$1.0 \times 10^{10}$ & 15 & 24 & 22.42 & 1.08 & 1000 & 868\\
\\
$1.0 \times 10^{11}$ & 5 & 11 & 22.68 & 2.00 & 1000 & 946\\
$1.0 \times 10^{11}$ & 10 & 11 & 29.34 & 2.15 & 1000 & 826\\
$1.0 \times 10^{11}$ & 15 & 11 & 33.51 & 2.15 & 1000 & 828\\
\\
$1.5 \times 10^{12}$ & 5 & 4.6 & 26.75 & 2.75 & 1000 & 960\\
$1.5 \times 10^{12}$ & 10 & 4.6 & 38.21 & 3.52 & 1000 & 614\\
$1.5 \times 10^{12}$ & 15 & 4.6 & 46.69 & 3.94 & 1000 & 483\\
\\
$1.5 \times 10^{12}$ & 5 & 4.6 & 26.75 & 2.75 & 16000 & 3749\\
$1.5 \times 10^{12}$ & 10 & 4.6 & 38.21 & 3.52 & 16000 & 2746\\
$1.5 \times 10^{12}$ & 15 & 4.6 & 46.69 & 3.94 & 16000 & 2152\\
\\
$1.0 \times 10^{13}$ & 5 & 2.4 & 28.45 & 3.10 & 1000 & 825\\
$1.0 \times 10^{13}$ & 10 & 2.4 & 42.60 & 4.34 & 1000 & 490\\
$1.0 \times 10^{13}$ & 15 & 2.4 & 54.02 & 5.19 & 1000 & 324\\
\\
$1.0 \times 10^{14}$ & 5 & 1.0 & 29.62 & 3.36 & 1000 & 587\\
$1.0 \times 10^{14}$ & 10 & 1.0 & 45.90 & 5.03 & 1000 & 314\\
$1.0 \times 10^{14}$ & 15 & 1.0 & 59.99 & 6.37 & 1000 & 190\\
\end{tabular}
}
\caption{
Disruption times -- defined as the time after which the cluster has lost half 
of its initial mass -- for clusters orbiting in NFW halos of different halo 
mass and concentration. We also show the percentage of the virial radius at 
which the cluster is orbiting at $8.5\,$kpc, the circular velocity and the 
gradient in the circular velocity at that radius, and the initial number of 
stars in the model.} \label{tb:disruption}
\end{center}
\end{table}

To quantify the disruption of a cluster we have defined the disruption time
scale as the time it takes for the cluster to lose half of its mass. These 
timescales are tabulated in Table \ref{tb:disruption}. Generally, clusters 
orbiting in more massive halos have shorter disruption times than those
orbiting in less massive halos. In massive halos the concentration has a 
large effect on the disruption time while in the lower mass ranges the 
disruption is independent of mass and concentration. Variations in this
regime are dominated by statistical effects. 

We can understand why this might be by considering the tidal field across a
cluster and determining how large it needs to be for the typical speeds of
stars in the cluster to exceed the velocity dispersion.
If $\sigma$ is the velocity dispersion of the cluster and $R_{\rm c}$
is its radius, then on a timescale $f \times R_{\rm c}/\sigma$ the net 
velocity of a star across the cluster is approximately $\partial a/\partial r
\times 2\,R_{\rm c} \times f \times R_{\rm c}/\sigma$. Here the factor $f$
will be greater than unity. If the test particle is to escape the cluster,  
$4\times f \times (V_{\rm c}/R)^2 \times R_{\rm c}^2/\sigma$ must be
comparable to $\sigma$, where $V_{\rm c}$ is the halo circular velocity at 
halo-centric radius $R$. This reduces to the condition that 
\begin{equation}
\sigma \sim 2 \sqrt{f} V_{\rm c} \left(\frac{R_{\rm c}}{R}\right)
\end{equation}
\noindent
For the star cluster mass we have considered, $\sigma \simeq 1\,$km/s
and $R_{\rm c}/R \simeq 0.001$, so $V_{\rm c} \sim 500\,$km/s if $f=1$, or in
other words in heavy halos clusters disrupt very efficiently. For less massive
halos, we require $f \sim 100$, and so internal processes rather than the
external tidal field will be important in driving the mass loss rate.\\

\begin{figure}
\begin{center}
\includegraphics[width=9cm]{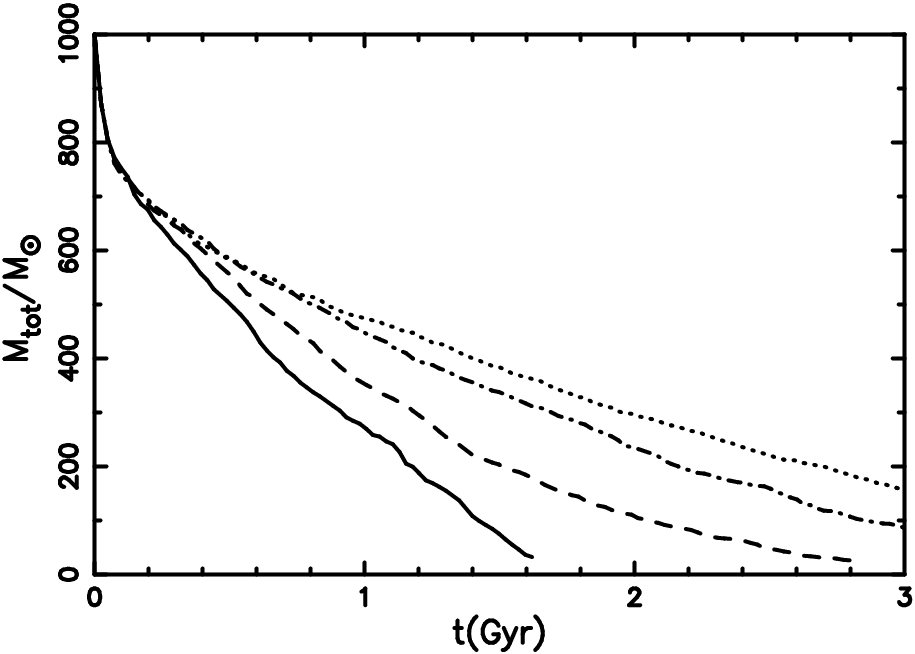}
\caption{The mass loss for the standard NFW model ($M_{\rm vir} = 1.5 \times
  10^{12}\,{\rm M_{\odot}}$, $c=10$) at different orbital distances: $4.5\,$kpc (solid),
  $8.5\,$kpc (dashed), $17\,$kpc (dashed-dotted) and $34\,$kpc
  (dotted). All models start with $N=1000$.} \label{fig:massNIII12}
\end{center}
\end{figure}

The model with virial halo mass of $1.5 \times 10^{12}\,{\rm M_{\odot}}$ and
concentration $c=10$ resembles closest a model for our own Milky Way halo and
from now on we will call it the standard NFW model. We ran $N=16\,000$
simulations for the models with this halo mass and the three different
concentrations. The results are shown in Table \ref{tb:disruption} 
for comparison. The tidal radius for these larger models is $40\,$pc, as given
by equation \ref{eq:king}. We see the same general trend of decreasing disruption time with
increasing concentration as seen for the $N = 1000$ models.
It has been shown previously by \citet{baumgardt2001} that the disruption time
of a star cluster can be expected to scale by the initial half-mass relaxation 
timescale and an $N$ dependent factor of $\left( \log \left( 0.11 N \right) / N \right)^{1/4}$ 
\citep[see also][]{trenti2007}. 
This is based on simulations of star clusters within a standard Galactic tidal field.
The disruption timescales of our $N = 16\,000$ models exceed those of
their $N = 1000$ counterparts by a factor of $4 - 4.5$ while the
half-mass relaxation timescale is a factor of two greater on average
in the larger models. Thus we do not agree with the \citet{baumgardt2001}
scaling relation by almost a factor of three. However, the use of both a tidal 
cut-off and a non-standard tidal field will certainly play a role and we will attempt 
to quantify this in future work after performing simulations for an expanded range of $N$.
 
For the standard NFW model we also looked at the effects of orbital
distance. Figure \ref{fig:massNIII12} shows the mass evolution for this standard model at four
different orbital distances. A smaller orbital radius leads to faster disruption.  
Note that the tidal radius remains fixed at $16\,$pc for these models.

\begin{figure}
\begin{center}
\includegraphics[width=9cm]{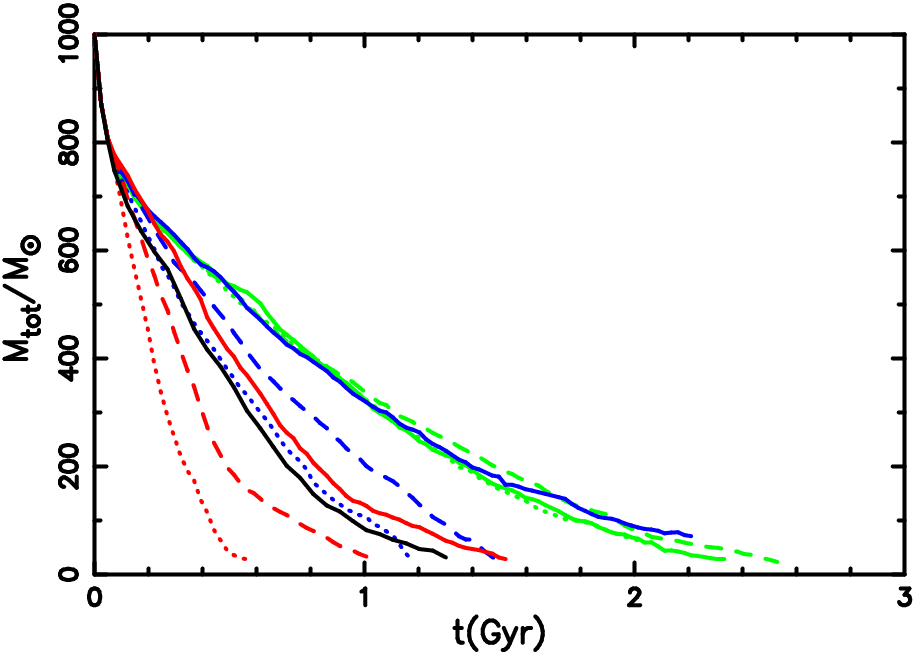}
\caption{The total mass evolution for a cluster at an orbital distance of
  $8.5\,$kpc within a full Galactic model (Model 5). Labels are as in Figure 
  \ref{fig:massNIII8} except the solid black line now represents Model 2, with
  the tidal field based on the Oort constants. All models start with $N=1000$} 
\label{fig:massNIV8}
\end{center}
\end{figure}

\subsection{The Three-Component Galaxy -- Bulge, Disk \& Halo}
In Model 5 we added the two-component model Galaxy (Model 3) to the NFW models
(Model 4), producing a three-component model consisting of bulge, disk and
dark matter halo. Figure \ref{fig:massNIV8} shows the mass evolution for this
model. The presence of the bulge and disk results in more rapid mass loss and
consequently a shorter
disruption time for the cluster. This is to be expected because of the
addition of mass. The relative effect is greatest on the lowest mass halos
because the size of the stellar component of the Galaxy is larger
relative to the mass of their halo than is the case for larger halos. Comparing this to Model 2
(external field based on the local Oort constants) we see that it can be well 
reproduced by representing the galaxy as a disk, bulge and dark matter halo
with a mass of $1.5 \times 10^{12}\,{\rm M_{ \odot}}$ and concentration c =
15. The enclosed mass of the dark matter halo at an orbital distance of
$8.5\,$kpc is $9.4 \times 10^{10}\,{\rm M_{\odot}}$. This is of order 20\% larger than
what one infers from the Oort constants.
All the models so far have assumed orbits in the disk. However we have looked
at a model whose orbital is perpendicular to the plane of the disk; this resulted 
in an enhanced disruption rate because of ``disk shocking": the cluster
experiences a sudden change in the potential as it goes through the disk 
resulting in faster mass loss. 

\section{Summary}
\label{summary}

We have examined the importance the presence of a
dark matter halo has on the rate at which a star cluster orbiting in its tidal
field loses mass. For our halo model we adopted the NFW profile, which
provides a good description of the mass profiles of dark matter halos forming
in cosmological simulations. The NFW profile can be characterised by a single 
parameter, its concentration, which provides a measure of how rapidly the
logarithmic slope changes with radius and consequently how rapidly the radial 
force changes with radius. As expected, star clusters orbiting at a fixed
physical radius lose mass at a higher rate in more massive halos and in more 
concentrated halos. 

We also investigated a three-component model of the Galactic tidal field,
consisting of a stellar bulge and disk and a NFW halo and compared the mass
loss rates of star clusters orbiting in this potential against mass loss rates
of clusters orbiting in a standard Galactic tidal field (SGTF). The SGTF is
based on empirical estimates of the Galactic tidal field in the Solar
neighbourhood based on stellar motions, and as such provides a measure of
total mass within the Solar radius, implicitly including our Galaxy's dark
matter halo. We showed that star clusters orbiting in three-component models lose
mass at a comparable rate to those orbiting in SGTF models if the halo mass is
$1.5 \times 10^{12}\,\rm M_{\odot}$ and concentration $c=15$. These numbers are
reasonable and consistent with what one would expect for Galactic-type dark
matter halos in cosmological simulations. By building multi-component models
of galactic potentials, one can explore many questions of cluster evolution
that have not necessarily been possible before (such as the disruption
timescales of clusters on eccentric orbits).

Our results are interesting because they highlight the importance of
including an underlying dark matter halo in $N$-body simulations of long-term globular
cluster evolution. Previous studies have neglected to do this, but as we have
shown, the halo can have a profound effect on the lifetime of the star
cluster. However, in the centre of the galaxy baryons will dominate the
mass contribution.
In future work we will also look at the effect of the halo mass on the 
internal properties of star clusters as they evolve. This is an important 
step in the spirit of the MODEST collaboration \citep{sills2003} to increase 
the realism of star cluster simulations. 
Of course, our calculation is idealised; we assume a static NFW halo 
-- one that does not evolve over time -- and our star clusters remain on a
circular orbit at a fixed radius over a Hubble time. As we have noted in the 
introduction, we would expect a realistic halo to have had an active and at 
times violent mass assembly history, which should be an important additional
driver of mass loss in our clusters. Furthermore, it is unclear
what effect the assembly of the galaxy would have on the structure of the
underlying dark matter halo. Nevertheless, our general conclusions hold and
the effect of non-circular orbits and an evolving halo will be qualified as we 
continue our study.

Our results raise a number of interesting questions.
The favoured cosmological model predicts that galaxies should live
in cuspy dark matter halos \citep{springel2006}. Previous studies of globular cluster evolution
have been of clusters that evolve in non-cuspy potentials. Inferences have
been drawn from these simulations of general processes such as core-collapse
and mass-segregation and observations of globular clusters, which the simulations
seem to do a good job of reproducing. We do not appear to need
cuspy dark matter halos to produce these these trends in simulated globular
clusters that are consistent with observational data, so does this imply that
cuspy dark matter halos are not needed at all? This is not straightforward
to answer, because it is not understood how the assembly of the galaxy has
affected its dark matter halo, and whether this has effectively wiped out the
cusp (as would appear necessary from rotation curve studies of dark matter
dominated galaxies). However, if cusps are robust in the presence of
the growing galaxy, then it may have implications for our understanding of
globular cluster evolution.

More speculatively, the enhanced rate of disruption of clusters in cuspy
halos has interesting implications for their use as probes of galaxy formation.
Understanding how efficiently globular clusters form should provide us with
insight into the efficiency of galaxy formation at high redshifts. The number
of old globular clusters today could tell us how plentiful the sites of globular
cluster formation were in the past, which in turn tells us about the efficiency
with which gas cooled and formed molecular clouds. These sites may be
regulated by the location (in the disks of gas-rich proto-galaxies, in mergers
between gas-rich proto-galaxies) or the ambient radiation field (cosmological
reionisation). If globular clusters are disrupted more efficiently in cuspy
halos, then we could be missing a large fraction of globular clusters thus
misinterpreting observations and drawing even more uncertain conclusions.

These are interesting questions and ones that we aim to revisit in coming papers.

\section*{Acknowledgements}

CP is supported by the STFC rolling grant for theoretical astrophysics at
University of Leicester.

\end{document}